\documentclass[journal,twoside,web]{IEEEtran}
\usepackage{cite}
\usepackage{amsmath,amssymb,amsfonts}
\usepackage{mathtools,amsthm}
\usepackage[bitstream-charter]{mathdesign}
\usepackage{mathalfa}
\usepackage{color}
\usepackage{bbm}
\usepackage{thmtools}
\usepackage{enumitem}
\usepackage{dsfont}
\usepackage{hyperref}
\newtheorem{lem}{Lemma}
\newtheorem{prop}{Proposition}

\renewenvironment{proof}{\vspace{.1cm}\noindent{\sc Proof.}\hspace{0.10cm}\,\,}{$\hfill\Box$\vspace{.1cm}}
\newcommand{\cE}{\mathcal{E}}
\newcommand{\cG}{\mathcal{G}}
\newcommand{\cJ}{\mathcal{J}}
\newcommand{\cK}{\mathcal{K}}
\newcommand{\cS}{\mathcal{S}}
\newcommand{\cU}{\mathcal{U}}
\newcommand{\cV}{\mathcal{V}}
\newcommand{\cW}{\mathcal{W}}
\let\norm\undefined
\newcommand{\bracket}[1]{\left( #1 \right)}
\newcommand{\norm}[1]{\left\| #1 \right\|}
\newcommand{\BR}[1]{\ensuremath{\left\lbrace #1 \right\rbrace}}

\newcommand{\diag}[1]{\ensuremath\text{diag}\bracket{#1}}
\newcommand{\bbm}[1]{\left[\begin{matrix} #1 \end{matrix}\right]}
\newcommand{\sbm}[1]{\left[\begin{smallmatrix} #1
	\end{smallmatrix}\right]}

\DeclareMathOperator*{\maxxx}{max}
\newcommand{\tmax}{\rm{max}}
\newcommand{\tmin}{\rm{min}}
\newcommand{\B}{\mathbb{B}}
\newcommand{\R}{\mathbb{R}}
\DeclareMathOperator*{\argmin}{arg\,min}
\def\revised{\color{black}}
\def\BibTeX{{\rm B\kern-.05em{\sc i\kern-.025em b}\kern-.08em
    T\kern-.1667em\lower.7ex\hbox{E}\kern-.125emX}}

\begin{document}
\title{Cooperative Nearest-Neighbor Control of Multi-Agent Systems: Consensus and Formation Control Problems}
\author{Muhammad Zaki Almuzakki 
and Bayu Jayawardhana, \IEEEmembership{Senior Member, IEEE}
\thanks{The work of Muhammad Zaki Almuzakki is  supported by \emph{Lembaga Pengelola Dana Pendidikan Republik Indonesia} (LPDP-RI) under contract No. PRJ-851/LPDP.3/2016. {\it (Corresponding author: Muhammad Zaki Almuzakki)}}
\thanks{Muhammad Zaki Almuzakki is with the Department of Computer Science, Faculty of Science and Computer,
        Universitas Pertamina, Jakarta, Indonesia
        (email: m.z.almuzakki@universitaspertamina.ac.id).}
\thanks{The authors are with the Engineering and Technology Institute Groningen, Faculty of Science and Engineering,  University of Groningen, The Netherlands
        (emails: \{m.z.almuzakki; b.jayawardhana\}@rug.nl).}
}

\maketitle
\thispagestyle{empty}

\begin{abstract}
This letter studies the problem of cooperative nearest-neighbor control of multi-agent systems where each agent can only realize a finite set of control points. Under the assumption that the underlying graph representing the communication network between agents is connected and the interior of the convex hull of all finite actions of each agent contains the zero element, consensus or distance-based formation problems can practically be stabilized by means of nearest-neighbor control approach combined with the well-known consensus control or distributed formation control laws, respectively. Furthermore, we provide the convergence bound for each corresponding error vector which can be computed based on the information of individual agent's finite control points. Finally, we show Monte Carlo numerical simulations that confirm our analysis.
\end{abstract}


\section{Introduction}
\label{sec:introduction}
\IEEEPARstart{T}{he} consensus (rendezvous/agreement) and formation control problems are two prototypical cooperative control problems in multi-agent systems (MAS). For systems with continuous input space, 
the problems of designing control laws to achieve consensus or to maintain a formation shape have been well-studied in the literature, for example \cite{Moreau2004, Ren2005, LI2018144,OlfatiSaber2006, oh2015survey}, among many others. However, practical implementation of {\revised MAS' control designs} may have to deal with physical constraints in the actuators, sensors and mechanisms, or with information constraints in the communication channel. Such constraints may be encountered due to the limitation of digital communication \cite{Carli2009,PARK2022105160} or due to the limitation of the mechanical design of the system such as the use of fixed set of discrete actuation systems in Ocean Grazer wave energy converter \cite{BarradasBerglind2016,Wei2018}. Designs, analysis, and numerical implementation of control laws for such networked systems have also received considerable attention in the literature, see for example \cite{Marcantoni2023, Zhang2013, dePersis2012,Jafarian2015}.

The temporal and spatial discretization of inputs, states and outputs of networked control systems are typically done via quantization operator. 
{\revised There are three classes of quantizers that are typically used in the literature, namely, uniform, asymmetric, and logarithmic quantizers} \cite{Wei2019}. The application and analysis of cooperative control with quantizers have been studied, for instance, in \cite{Marcantoni2023, Zhang2013, dePersis2012,Jafarian2015,Wei2019,cortes2006finite,Ceragioli2015,Sun2016,XIAO201759}. However, when minimality requirement is imposed on the number of control input points or on the number of symbols in the communication channel, the design and analysis tools using  aforementioned quantizers can no longer be used to address this problem. An example of such case is mechanical systems with finite discrete actuation points as in \cite{BarradasBerglind2016,Wei2018}.

In \cite{JAYAWARDHANA2019460,Almuzakki2022}, these quantization operators are considered as nearest-neighbor operators that map the input value to the available points in a given discrete set $\mathcal U$, which can have a finite or infinite number of members. The authors study the use of $\mathcal U$ with minimal cardinality such that the closed-loop systems are practically stable. Particularly, it is shown that for a generic class of $m$-dimensional passive systems having proper storage function and satisfying the nonlinear large-time initial-state norm observablility condition\footnote{We refer interested readers to  \cite{hespanha2005nonlinear} for a reference to the notion of nonlinear norm observability.}, it can be practically stabilized using only $m+2$ control actions. {
\color{black} As a comparison, using the $q$-ary quantizers\footnote{In this case, binary quantizer is given by $q=2$ and ternary quantizer corresponds to $q=3$.} \cite{dePersis2012,Jafarian2015,DEPERSIS2009602}, where $q$ input values per input dimension are defined, the stabilization of the systems requires $\mathcal U$ whose cardinality is $q^m$ (or $q^m+1$ if the zero element is not in the range of the $q$-ary quantizers).} 

In this letter, we present the application of nearest-neighbor control to the cooperative control of multi-agent systems. We study the combination of the nearest-neighbor approach studied in \cite{JAYAWARDHANA2019460,Almuzakki2022} and the standard distributed continuous control laws for multi agent-cooperation as in \cite{dePersis2012,oh2015survey,Marcantoni2023}. 
Specifically, we study nearest-neighbor distributed control for consensus and distance-based formation control problems.
{\revised We emphasize that the notion of {\it nearest-neighbor control} is consistent with the prior work in [19]-[20] and it is not related to the notion of neighbors in the graph of multi-agent systems.}
We show the practical stability property of the closed-loop system where the usual consensus and distance-based formation Lyapunov function are used in the analysis. We present the upper bound of the practical stability of the consensus or formation error that can be computed based on the local bound from each individual $\mathcal U_i$ at each agent. 

The rest of the letter is organized as follows. Some notations and preliminaries on continuous consensus and distance-based formation control design in addition to the relevant properties of the nearest-neighbor operator are presented in Section~\ref{sec:preliminaries}. In Section~\ref{sec:main}, we present our main results on the nearest-neighbor consensus and distance-based formation control laws along with the upper bound analysis on the practical stability of the error. In Section~\ref{sec:simulation}, we show numerical analysis using Monte Carlo simulations that show the validity of our main results. Finally, the letter is concluded with conclusions
in Section~\ref{sec:conclusion}.

\section{Preliminaries and Problem Formulation}\label{sec:preliminaries}

{\bf {\revised Notation}:} For a vector in $\R^n$, or a matrix in $\R^{m\times n}$, we denote the Euclidean norm and the corresponding induced norm by $\| \cdot \|$. {\color{black} The direct sum of two vector spaces is denoted by $\oplus$. The Kronecker product of two matrices is denoted by $\otimes$. For a linear mapping $T(x) = Ax$, we denote the kernel and image of $T$ by $\text{Ker}(A)$ and $\text{Im}(A)$, respectively.}
For any point $c\in\R^n$, the set $\mathbb B_\epsilon(c) \subset \R^n$ is defined as, $\mathbb B_\epsilon(c) :=\{\xi\in \R^n | \|\xi-c\|\leq \epsilon\}$. For simplicity, we write $\mathbb B_\epsilon(0)$ as $\mathbb B_\epsilon$. {\color{black} Furthermore, we write $\mathbb B_\epsilon\subseteq\R^n$ as $\mathbb B_\epsilon^n$.} The inner product of two vectors $\mu,\nu\in \R^m$ is denoted by $\langle \mu, \nu \rangle$. For a given set $\mathcal{S}\subset\R^m$, and a vector $\mu \in \R^m$, we let $\langle \mu, \cS \rangle := \{ \langle \mu, \nu \rangle \, \vert \, \nu \in \cS \}$. For a discrete set $\cU$, its cardinality is denoted by $\text{card}(\cU)$. The convex hull of vertices from a discrete set $\cU$ is denoted by $\text{conv}(\cU)$. The interior of a set $S\subset\R^n$ is denoted by $\text{int}\bracket{S}$. 
For a countable set $\mathcal{S}\subset\R^m${\revised , the} Voronoi cell of a point $s\in\mathcal{S}$ is defined by $V_\mathcal{S}(s) := \BR{x\in\R^m\ |\ \|x-s\|\leq\|x-v\|,\ \forall v\in\mathcal{S}\setminus\{s\} }$.
{\revised
For a discontinuous map $F:\R^n\to\R^n$, the Krasovskii regularization of $F$ is the set-valued map defined by $\cK(F(x)):=\bigcap_{\delta>0}\text{conv}(F(x+\B_\delta))$.}\\[0.25ex]

As discussed in the Introduction, we will study the use of nearest neighbor control for solving two multi-agent problems of consensus and formation control. In this regards, we consider an undirected graph $\mathcal G = (\mathcal V, \mathcal E)$ for describing the network topology, where $\mathcal V$ is the set of $N$ agents and $\mathcal E\subset \mathcal V\times \mathcal V$ is a set of $M$ edges that define the neighboring pairs. Moreover we assume that the graph $\mathcal G$ is connected. For every edge $k$ in 
{\revised $\mathcal G$,} 
we can associate one node by a positive sign and the pairing node by a negative sign. Correspondingly, the incidence matrix $B\in\R^{N\times M}$ can be defined by 
\[
b_{i,k} = \left\{\begin{array}{ll}+1 & \text{if node }i\text{ has the positive sign in edge }k \\ -1 & \text{if node }i\text{ has the negative sign in edge }k\\0 & \text{otherwise}\end{array}\right.
\]
Using $B$, the Laplacian matrix $L$ is given by $L=BB^\top$ whose kernel, by the connectedness of $\mathcal G$, is spanned by $\mathds{1}_{N}$. 

\subsection{Multi-Agent Consensus}\label{sec:cont_consensus}

For every agent $i$ in $\mathcal G$, it is described by 
\begin{equation}\label{eq:main}
    \dot x_i = u_i.
\end{equation}
where $x_i(t)\in\R^m$ and $u_i(t)\in\R^m$ denote the state and input variables, respectively.  
The distributed consensus control problem is related to the design of distributed control law $u_i$ for each agent based on the information from the neighboring agents so that all agents converge to a consensus point. The well-known control law  $u = -(L\otimes I_m)x$ solves this problem, where it can be shown that by using the consensus Lyapunov function $V(x)=\frac{1}{2}x^\top \bracket{L\otimes I_m} x$, $\lim_{t\to \infty}\|x_i(t)-\bar x\| = 0$ for all $i$ and $\bar x=\frac{1}{N}\sum_ix(0)\in\R^m$. We define the consensus manifold $E$ where all agents agree with each other by {\revised $E:=\{\bar x \in \R^{mN} | \bar x = \bar x_1 = \bar x_2 = \ldots = \bar x_N\}$}. 

{\color{black}
The stability of the closed-loop system is, in fact, carried out by introducing the relative position variable 
\begin{equation}
    z_k=\begin{cases}
        x_i-x_j & \text{if node $i$ is the positive end of edge $k$},\\
        x_j-x_i & \text{if node $i$ is the negative end of edge $k$},
    \end{cases}
\end{equation}
and we write its compact form as $z=(B^\top\otimes I_m) x$. The closed-loop system of the consensus problem is then expressed as
\begin{equation}\label{eq:closedLoop_consensus}
    \dot z = -(B^\top B\otimes I_m)z
\end{equation}
and the consensus Lyapunov function becomes $V(z)=\frac{1}{2}z^\top z$ so that stability can then be shown by using LaSalle's invariance principle. That is, $z\to 0$ as $t\to\infty$.
}

The generalization of the result to the case, where binary and ternary quantizers are used, can be found in \cite{DEPERSIS2009602,dePersis2012,Jafarian2015}.

\subsection{Distance-Based Multi-Agent Formation Control}\label{sec:formation_cont}

Consider the same set of $n$ agents as described in section~\ref{sec:cont_consensus}. The distributed distance-based formation control problem is, in principal, similar to the control design for consensus problem. The main difference is that in the asymptote, 
all agents must converge to a prescribed formation shape represented by the graph $\cG=(\cV,\cE)$ and the given desired distance between connected agents. For given desired distance $d_{k}$ associated to the relative position $z_k$, $k=1,\dots,M$, the well-known control law $u=-(B\otimes I_m)D_z e$ where $D_z$ takes the form of the block-diagonal matrix $D_z:=\diag{z}\in\R^{Mm\times M}$ and $e$ is the desired error vector defined by
\begin{equation}\label{eq:error_vec}
e=\bbm{\norm{z_1}^2-d_1^2,&\cdots,& \norm{z_{M}}^2-d_{M}^2}^\top
\end{equation}
solves the distance-based distributed formation control.

The stability of above distributed formation control problem can be analyzed by considering the dynamics of the closed-loop autonomous multi-agent system given by
\begin{align}
    \dot z&=(B^\top\otimes I_m) \dot x = -(B^\top B\otimes I_m) D_z e\label{eq:cl_z_cont}\\
    \dot e&=D_z^\top \dot z = -D_z^\top(B^\top B\otimes I_m)D_z e.\label{eq:cl_e_cont}
\end{align}
Using the usual distance-based formation Lyapunov function $J(e)=\frac{1}{4}\langle e, e \rangle$, the local exponential convergence of $e$ to zero can be shown, which means that 
$\|z_k(t)\|\to d_k$ locally and exponentially as $t\to\infty$.

\subsection{Nearest-Neighbor Map}

\begin{enumerate}
    \item[(A1)] For a given set $\mathcal U:= \{0, u_1, u_2, \ldots, u_p\}$, there exists an index set $\mathcal I\subset \{1,\ldots,p\}$ such that the set $\mathcal V:=\{u_i\}_{i\in \mathcal I}\subset \mathcal U$ defines the vertices of a convex polytope satisfying, $0\in\text{int}\bracket{\text{conv}\bracket{\mathcal V}}$.
\end{enumerate}

\begin{lem}[{ \cite[Lemma 1]{Almuzakki2022} }]\label{lemma:1}
Consider a discrete set $\cU \subset \R^m$ that satisfies (A1). Then, there exists $\delta > 0$ such that
\begin{equation}\label{eq:bndVoronoi}
V_{\cU} (0) \subseteq \B_\delta,
\end{equation}
where $V_{\cU}$ is the Voronoi cell of $\cU$ as defined before. In other words, 
the following implication holds for each $\eta \in \R^m$
\begin{equation}\label{eq:defCv}
    \| \eta\| > \delta \Rightarrow \ \exists \ u_i \in \cU \text{ s.t. } \|\eta - u_i \|< \| \eta\|.
\end{equation}
\end{lem}

{\revised
We define the nearest-neighbor mapping $\phi_i:\R^m \rightrightarrows \cU_i$ as
\begin{equation}\label{phi_eq}
\phi_i(\eta) := \argmin_{v\in \mathcal{U}_i}\left\{{\|v-\eta\|}\right\}.
\end{equation}

\begin{lem}\cite{Almuzakki2022}\label{lem:phi_bndIneq}
    Consider the nearest-neighbor mapping $\phi_i$ given in \eqref{phi_eq} and a discrete set $\mathcal U_i:=\{0,u_{1},u_{2},\dots,u_{p}\}$ satisfying (A1). For a fixed $y\in\R^m$, let $\phi_i(-y)=\{u_j\}_{j\in\cJ}$ for some index set $\cJ\subset\{1,\dots,p\}$. Then the inequality
    \begin{equation}\label{eq:bndUpLowInnProd}
    -\|u_j\| \cdot \|y\| \le \langle u_j,y \rangle \le -\frac{1}{2} \|u_j\|^2
    \end{equation}
    holds for all $j\in\cJ$.
\end{lem}

We refer to \cite{Almuzakki2022} for the proof of Lemma~\ref{lem:phi_bndIneq}. By the definition of $\phi_i$, the inequality $\|u_j+y\|^2\le \|u_k+y\|^2$ holds for $j\in \cJ$ and $k\in\{0,1,\ldots, p\}$. By noting that $\|u_j+y\|^2 = \langle u_j+y,u_j+y\rangle = \|u_j\|^2 + 2\langle u_j,y\rangle + \|y\|^2$ and fixing $u_k=0$, we have that $\langle u_j,y \rangle \le -\frac{1}{2} \| u_j\|^2$. Moreover $\langle u_j,y \rangle\ge-\norm{u_j}\norm{y}$. Hence, the inequality \eqref{eq:bndUpLowInnProd} holds for every $y\in\R^m$. 
}

\section{Main Results}\label{sec:main}
{\color{black}

Prior to presenting the main results, we need the following technical lemma, which establishes the relationship between a ball in the range of $(B\otimes I_m)z$ and a ball of the same radius in $z$. It is used later to get an upperbound on the practical stability of the consensus or formation error. 

\begin{lem}\label{lem:main}
    Consider an undirected and connected graph $\mathcal G = (\mathcal V, \mathcal E)$. 
    Let $x_i\in\R^m,\ i=1,\dots,N,$ be the {\revised state variable} of {\revised the $i$-th agent as in \eqref{eq:main}} 
    and define $z=(B^\top\otimes I_m)x\in\R^{Mm}$. {\revised If both $(B\otimes I_m)z \in \mathbb B^{Nm}_\delta$ and $z\in \text{Im}(B^\top\otimes I_m)$ hold then $z\in \mathbb B^{Mm}_{\delta}$.} 
\end{lem}

\begin{proof}
    Firstly, by defining the space $\Omega:=\text{Ker}(B\otimes I_m) \oplus \Big(\text{Im}(B^\top\otimes I_m)\cap\mathbb B^{Mm}_{\delta}\Big)$, if $z\in \Omega$ then $(B\otimes I_m)z\in \text{Im}(B\otimes I_m)\cap\mathbb B^{Nm}_{m\|B\|\delta}$ (which is a superset ball that contains $B^{Nm}_{\delta}$).  
Since $z=(B^\top\otimes I_m)x$, {\revised it necessarily holds that} $z\in \text{Im}(B^\top\otimes I_m)$. Combining this with $z\in\Omega$, 
$\|(B\otimes I_m)z\|\leq \delta$ implies that $z\in \Omega \cap \text{Im}(B^\top\otimes I_m)$. 
Since the non-zero elements of $B$ are either $1$ or $-1$ and since the graph is connected, it follows that for all $z\in\Omega \cap \text{Im}(B^\top\otimes I_m)$, we have $\|z\|\leq \|(B\otimes I_m)z\| \leq m\|B\|\delta$. Hence, for all $z\in \Omega\cap \text{Im}(B^\top\otimes I_m)$, if $\|(B\otimes I_m)z\|\leq \delta$ then $\|z\|\leq \delta$. Moreover, by definition $\text{Ker}(B)\cap \text{Im}(B^\top) = \emptyset$, so that  $z\in \Big(\text{Ker}(B\otimes I_m)\cap \text{Im}(B^\top\otimes I_m) \Big) \oplus \Big(\text{Im}(B^\top\otimes I_m)\cap\mathbb B^{Mm}_{\delta}\Big) = \text{Im}(B^\top\otimes I_m)\cap\mathbb B^{Mm}_{\delta}$. 
We can now conclude that if both $\|(B\otimes I_m)z\|\leq \delta$ and $z\in \text{Im}(B^\top\otimes I_m)$, then $\|z\|\leq \delta$. 
\end{proof}

}

\subsection{Consensus Protocol With Finite Set of Actions}\label{sec:consensus}

{\color{black}

In this subsection, we propose a nearest-neighbor input-quantization approach for solving the practical consensus problem. 
In this case, every agent $i\in\{1,\dots,n\}$ is given by a single-integrator dynamics \eqref{eq:main} and its control input takes value from a set of finite points $\mathcal U_i:=\{0,u_{i,1},u_{i,2},\dots,u_{i,{p_i}}\}$ satisfying (A1) along with their respective quantity $\delta_i$ satisfying \eqref{eq:defCv}. For this  problem, we propose a nearest-neighbor controller for consensus problem by assigning $u_i=\phi_i(-(L\otimes I_m) x)$ with $\phi_i$ as in \eqref{phi_eq}. The corresponding closed-loop system can be written as
\begin{equation}\label{eq:cl_nnc_x}
    \dot x = \Phi(-(L\otimes I_m) x)
\end{equation}
where $\Phi$ is understood agent-wise, i.e.
\begin{equation}\label{eq:Phi}
  \Phi(\eta) = \bbm{\phi_1(\eta_1)^\top,&\cdots,&\phi_n(\eta_n)^\top}^\top. 
\end{equation}
In the relative position coordinate, \eqref{eq:cl_nnc_x} can be rewritten as
\begin{equation}\label{eq:cl_nnc_z}
    \dot z = (B^\top \otimes I_m)\Phi(-(B\otimes I_m) z).
\end{equation}
The stability of \eqref{eq:cl_nnc_z} is shown in the following proposition.

}

\begin{prop}\label{prop:consensus}
For given sets of finite control points $\mathcal U_i:=\{0,u_{i,1},u_{i,2},\dots,u_{i,{p_i}}\},\ i=1,\dots,N,$ satisfying (A1) along with their respective Voronoi cell upper bound $\delta_i$ satisfying \eqref{eq:defCv},
consider the closed-loop MAS in \eqref{eq:cl_nnc_z}, where $\Phi$ is as in \eqref{eq:Phi}. {\revised Then for any initial condition $z(0)=z_0$, 
$z(t)\to\B_{\delta}$ as $t\to\infty$ where $\delta=\sum\limits_{i=1}^N \delta_i$.}
\end{prop}

\begin{proof}
{
\color{black}
As pursued in \cite{Almuzakki2022}, since $\Phi$ is a non-smooth mapping, we can embed the differential equation \eqref{eq:cl_nnc_z} into a regularized differential inclusion given by
\begin{equation}
    \dot z \in (B^\top\otimes I_m) \cK(\Phi(-(B\otimes I_m)z)).\label{eq:diff_incl_Z}
\end{equation}
Using the usual consensus Lyapunov function $V(z)=\frac{1}{2}z^\top z$, it follows that
{\revised 
\begin{align*}
    \dot V(z) &\in \langle (B\otimes I_m)z, \cK(\Phi(-(B\otimes I_m)z))\rangle\\
    &= \sum\limits_{i=1}^n \langle (b_i\otimes I_m) z,\cK(\phi_i(-(b_i\otimes I_m) z))\rangle\\
    &= \sum\limits_{i=1}^n \langle (b_i\otimes I_m)z,{\rm conv}(\cW_i^c)\rangle,
\end{align*}
}
where $b_i$ is the $i$-th row vector of the incidence matrix $B$ {\revised and $\cW_i^c:=\phi_i(-(b_i\otimes I_m) z)$}.
Following Lemma \ref{lem:phi_bndIneq}, it follows that for every $i\in\{1,\ldots,N\}$, we have that 
\begin{itemize}
    \item if {\revised $0\not\in\cW_i^c$}, then
    {\revised
    \begin{multline*}
        \langle (b_i\otimes I_m)z,{\rm conv}(\cW_i^c) \rangle\\ \subset[-\norm{u_i^{\tmax}}\norm{(b_i\otimes I_m)z}, -0.5 \norm{u_i^{\tmin}}^2]
    \end{multline*}
    where $\norm{u_i^{\tmax}}=\maxxx\limits_{w_i\in\cW_i^c}\norm{w_i}$ and $\norm{u_i^{\tmin}}=\min\limits_{w_i\in\cW_i^c}\norm{w_i}$;} 
    or else
    \item if {\revised ${0}=\cW_i^c$}, then
    {\revised
    \begin{equation*}
    \langle (b_i\otimes I_m)z,\ {\rm conv}(\cW_i^c) \rangle = \{0\}.
    \end{equation*}
    }
\end{itemize}

Hence, for any given time $t\geq 0$, whenever 
{\revised $-(b_i\otimes I_m)z(t)\notin \text{int}(V_{\cU_i}(0))$}
for some $i$, we have $\dot V(z(t))<0$, i.e., the Lyapunov function $V(z(t))$ is strictly decreasing. Otherwise $\dot V(z(t)) =  0$. This implies that all Krasovskii solutions of \eqref{eq:cl_nnc_z} converge to the invariant set 
{\revised $\Psi=\{z | -(b_i\otimes I_m)z \in \text{int}(V_{\cU_i}(0)), \,\forall i\}$}. In the set $\Psi$, for each $i=1,\dots,N$, it must be that $\| (b_i\otimes I_m)z \|  \le \delta_i$. Thus
    \[
    \|(B\otimes I_m)z\| \le \sum\limits_{i=1}^n \| (b_i\otimes I_m)z \| 
    \le \sum\limits_{i=1}^n \delta_i=\delta.
    \]

By using Lemma~\ref{lem:main} and since $\|(B\otimes I_m)z\|\le\delta$ and $z=(B^\top\otimes I_m)x$, we can conclude that $\|z\|\le\delta$. 
}

It has been shown above that the relative position coordinate $z$ converges to a ball with size relative to the finite sets of actions of all agents and the network topology. Consequently, all agents represented by position $x_i, i=1,\dots,N$ are said to reach consensus in the neighborhood of the consensus manifold $E$.
\end{proof}

\subsection{Distance-Based Formation With Finite Sets of Actions}

Consider a set of $n$ agents governed by the single integrator dynamics, where {\revised each agent} can take value only from a given set of finite points $\cU_i$ as in subsection \ref{sec:consensus}. Given a desired distance vector 
$d=\bbm{d_1& \cdots & d_{M}}^\top$ 
representing desired distance constraints that define the desired formation shape, where for each $k=1,\dots,M$, $d_k=d_{ij}$ is the desired distance between the $i$th and $j$th agent in the formation. For this problem, we propose the nearest-neighbor distance-based control law $u=\Phi(-(B\otimes I_m)D_z e)$ with $\Phi$ be as in \eqref{eq:Phi}, $D_z$ and $e$ be as described in subsection \ref{sec:formation_cont}. In this case, the closed-loop system represented by \eqref{eq:cl_z_cont} and \eqref{eq:cl_e_cont} becomes
\begin{align}
    \dot z&= (B^\top\otimes I_m)\Phi(-(B\otimes I_m)D_z e)\label{eq:cl_z_nnc}\\
    \dot e&=D_z^\top (B^\top\otimes I_m)\Phi(-(B\otimes I_m)D_z e).\label{eq:cl_e_nnc}
\end{align}
The stability of above system is analyzed in the following proposition.

\begin{prop}\label{prop:formation}
    For given sets of finite control points $\mathcal U_i:=\{0,u_{i,1},u_{i,2},\dots,u_{i,{p_i}}\},\ i=1,\dots,N,$ satisfying (A1) along with their respective Voronoi cell upper bound $\delta_i$ satisfying \eqref{eq:defCv}, 
consider the closed-loop MAS \eqref{eq:cl_z_nnc} and \eqref{eq:cl_e_nnc} where $\Phi$ is as in \eqref{eq:Phi}. Then {\color{black} for any initial condition $(z(0),e(0))$ in the neighborhood of the desired formation shape, there exists $\bar\delta>0$ such that $\dot{z}(t)\to 0$, $\dot{e}(t)\to 0$ and $e(t)\to \mathbb B_{\bar\delta}$.} 
\end{prop}

\begin{proof}
Similar to the proof of Proposition~\ref{prop:consensus}, since $\Phi$ is a non-smooth mapping, we consider instead the regularized differential inclusion of the closed-loop systems given by
\begin{align}
    \dot z&\in (B^\top\otimes I_m)\cK(\Phi(-(B\otimes I_m)D_z e))\label{eq:incl_cl_z_nnc}\\
    \dot e&\in D_z^\top (B^\top\otimes I_m)\cK(\Phi(-(B\otimes I_m)D_z e)).\label{eq:incl_cl_e_nnc}
\end{align}

Using the usual distance-based formation Lyapunov function $J(e)=\frac{1}{4}\langle e, e \rangle$, it follows that
\begin{align*}
    \dot J(e)&=\langle e, D_z^\top(B^\top\otimes I_m)\Phi(-(B\otimes I_m)D_z e)\rangle\\
    &= \langle (B\otimes I_m)D_z e, \Phi(-(B\otimes I_m)D_z e)\rangle\\
    &\in \Big\langle (B\otimes I_m)D_z e ,\cK(\Phi(-(B\otimes I_m)D_z e))\Big\rangle\\
    &= \sum\limits_{i=1}^n \Big\langle (b_i\otimes I_m)D_z e ,{\rm conv}(\cW_i^f)\Big\rangle,
\end{align*}
where $\cW_i^f:=\phi_i(-(b_i\otimes I_m)D_z e)$.
Following similar computation as before, for every $i\in\{1,\dots,N\}$, we have that
\begin{itemize}
    \item if $0\not\in\cW_i^f$, then
    \begin{multline*}
        \langle (b_i\otimes I_m)D_z e,{\rm conv}(\cW_i^f) \rangle\\ \subset \left[-\norm{u_i^{\tmax}}\norm{(b_i\otimes I_m)D_z e}, -0.5 \norm{u_i^{\tmin}}^2\right]
    \end{multline*}
    where $\norm{u_i^{\tmax}}=\maxxx\limits_{w_i\in\cW_i^f}\norm{w_i}$ and $\norm{u_i^{\tmin}}=\min\limits_{w_i\in\cW_i^f}\norm{w_i}$; else
    \item if $\{0\}=\cW_i^f$, then
    \begin{equation*}
    \langle (b_i\otimes I_m)D_z e,\ {\rm conv}(\cW_i^f) \rangle = \{0\}.
    \end{equation*}
\end{itemize}

Hence, at any given time $t\geq 0$, whenever 
{\revised
$-(b_i\otimes I_m)D_z e\notin \text{int}(V_{\cU_{i}}(0))$ 
}
for some $i$, we can conclude that the Lyapunov function $J(e(t))$ is strictly decreasing. Otherwise $\dot J(e(t)) =  0$. {\color{black} By the radially unboundedness of $J(e)$, this means that as $t\to\infty$, the error function $e$ converges to a ball $\B_{c_e}$ for some $c_e>0$. Moreover, since $\|z\|$ can be written as a continuous function of $e$, namely $\|z\|=\sqrt{\sum\limits_{k=1}^M |e_k + d_k^2|}$, we also have that $z\in\B_{c_z}$ for some $c_z>0$. } The boundedness of $e$ and $z$ implies that all Krasovskii solutions of the system \eqref{eq:incl_cl_z_nnc} and \eqref{eq:incl_cl_e_nnc} converge to the invariant set 
{\revised
$\Psi=\{(z,e) | -(b_i\otimes I_m)D_z e \in \text{int}(V_{\mathcal U_{i}}(0)), \,\forall i\}$ 
}
where the state $(z,e)$ remains stationary. 

{\color{black} For the rest of the proof, we analyze the bound of $e$ in the invariant set $\Psi$ so that we can obtain the ball size around the origin where the formation error state $e$ converges to.} By the definition of $\Psi$ above, it follows that 
\begin{align*}
    \| (b_i\otimes I_m)D_z e \| 
    \le \delta_i,
\end{align*}
holds for all $e\in \Psi$ and for all $i=1,\dots,n$. 
Hence we have that
\begin{align*}
    \| (B\otimes I_m)D_z e \|&\le \sum\limits_{i=1}^{n} \| (b_i\otimes I_m)D_z e \| \\ 
    &\le \sum\limits_{i=1}^{n} \delta_i=:\delta.
\end{align*}

Using the same argumentation as in the proof of Proposition \ref{prop:consensus}, we can conclude using Lemma \ref{lem:main} that both $\| (B\otimes I_m)D_z e \| \leq \delta$ and $D_z e\in \text{Im}(B^\top \otimes I_m)$ imply that $\|D_ze\|\leq \delta$. 
Note that 
\begin{align}
 \label{eq:norm_Dze}
\|D_ze\| & = \sqrt{e^\top D^\top_{z}D_{z} e} 
 = \sqrt{e^\top D_{\tilde z} e},
\end{align}
where {\revised $\tilde z = \sbm{\|z_1\|^2 &\cdots& \|z_M\|^2}^\top$}. We will now establish the local practical stability of the closed-loop systems for the error state $e$. Using the radially unbounded function $J(e(t))$ which is non-increasing as a function of $t$, $\|e(t)\|\leq \|e(0)\|$ for all $t\geq 0$. Let us initialize the agents in the neighborhood of the desired formation shape, so that 
$\|e(0)\|< \min\{d_i^2\} = c_1$. Thus, in this case, 
\begin{align*}
\|z(t)\|^2 & = \sum\limits_{k=1}^M |e_k(t) + d_k^2| 
\geq \sum\limits_{k=1}^M (d_k^2 - c_1)=c_2^2>0, 
\end{align*}
for all $t\geq 0$ and for some $c_2>0$. 
Combining this with \eqref{eq:norm_Dze}, 
we get
$ \|D_ze\|   = \sqrt{e^\top D_{\tilde z} e}
  \geq c_2\|e\|$. 
Hence we can conclude that in the invariant set $\Psi$, 
we have $\|e\|\leq \frac{1}{c_2}\|D_ze\|\leq \frac{\delta}{c_2}$. 
\end{proof}

\section{Numerical Simulations}\label{sec:simulation}

In this section, we provide numerical analysis to the proposed cooperative nearest-neighbor control of multi-agent systems, for both the consensus problem, as well as, the formation control problem. 

For the numerical analysis, we perform Monte-Carlo simulations with $1000$ samples of simulation with the following simulation setup: 
\begin{enumerate}
    \item for each simulation, the number of agents are generated randomly between 3 to 7 agents;
    \item the agents are initialized in equidistant circular positions with prescribed rigid \emph{communication} networks {\revised and then placed on the 2-dimensional Euclidean space with additional random numbers to the initial coordinates};
    \item {\revised each agent} can only realize motion in three distinct directions in the direction of the vertices of an equilateral triangle with fixed length or stay at their current position. The set of actions realizable by {\revised each agent} is described by
    \begin{align*}
        \cU_i&=\\
        &\delta_i\sbm{\cos(\theta_i) & -\sin(\theta_i)\\\sin(\theta_i) & \cos(\theta_i)}\left\{ \sbm{0\\0}, \sbm{\sin(0)\\\cos(0)}, \sbm{\sin(\frac{2\pi}{3})\\\cos(\frac{2\pi}{3})}, \sbm{\sin(\frac{4\pi}{3})\\\cos(\frac{4\pi}{3})} \right\}
    \end{align*}
    where $\delta_i$ is the {\revised smallest upper-bound} of Voronoi cell  
    {\revised satisfying Lemma~1}
    for {\revised each agent} $i=1,\dots,N$ as in \cite[Example 2]{Almuzakki2022} and $\theta_i$ is the randomized rotation angle within {\revised the interval $[0,2\pi)$;}
    \item for each simulation, the corresponding $\delta_i$ of each agent is chosen randomly so that $\sum_i \delta_i = 1$, i.e.\ the maximum error bound is 1; and  
    \item the results are processed to obtain the 95\% confidence interval statistics for the error vectors, which is the vector $z$ for the consensus problem and the vector $e$ for the formation control problem. We also analyze their minimum and maximum trajectories.
\end{enumerate}

\begin{figure}[hbtp!]
    \centering
    \includegraphics[width=0.95\linewidth]{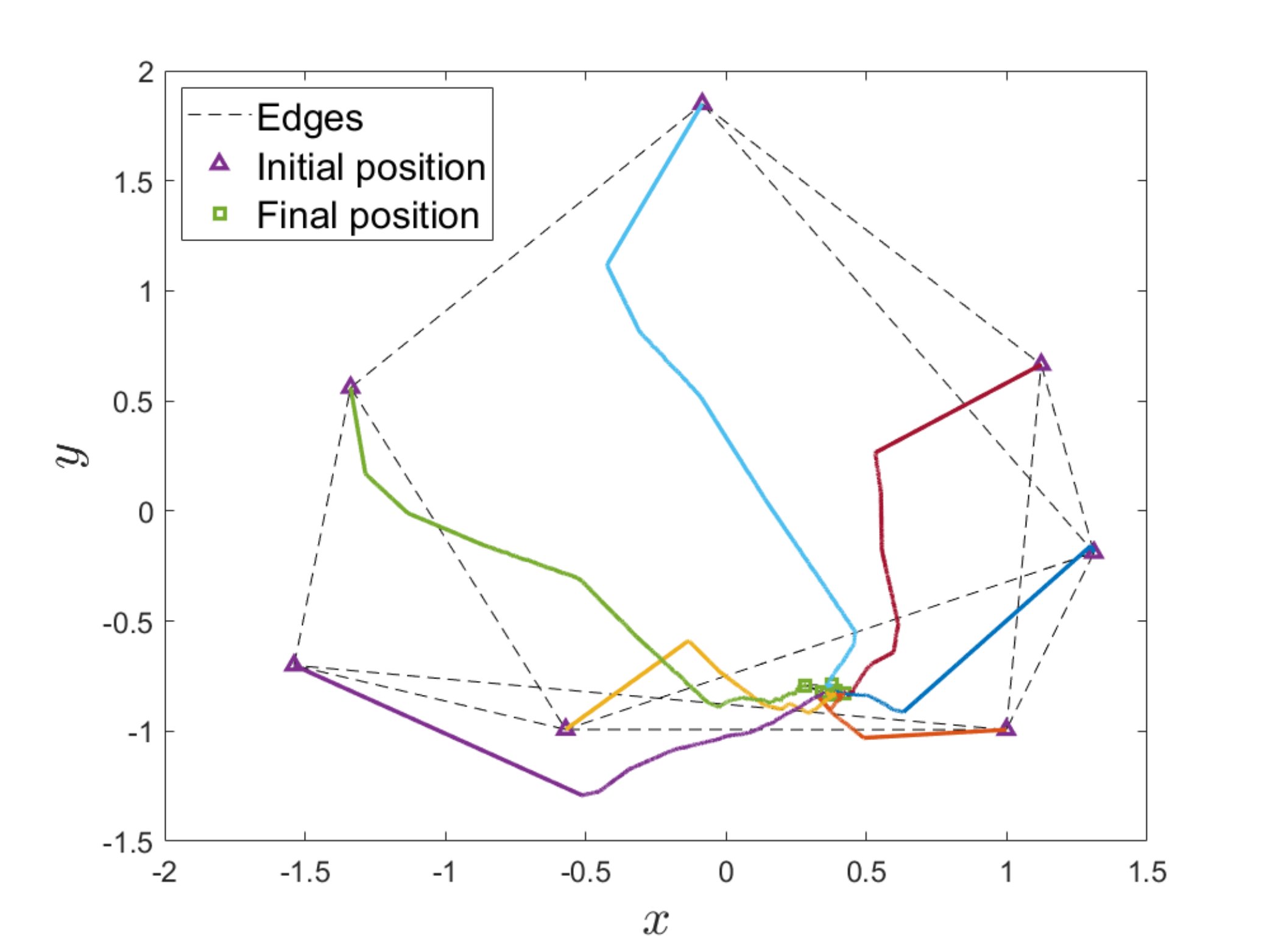}
    \caption{An example of consensus mechanism of a system with seven agents communicating over a rigid network where series of actions are chosen by means of nearest-neighbor consensus protocol. This example is taken from one of the 1000 random simulations.}
    \label{fig:consensus_pos_evol}
\end{figure}

\begin{figure}[hbtp!]
    \centering
    \includegraphics[width=0.95\linewidth]{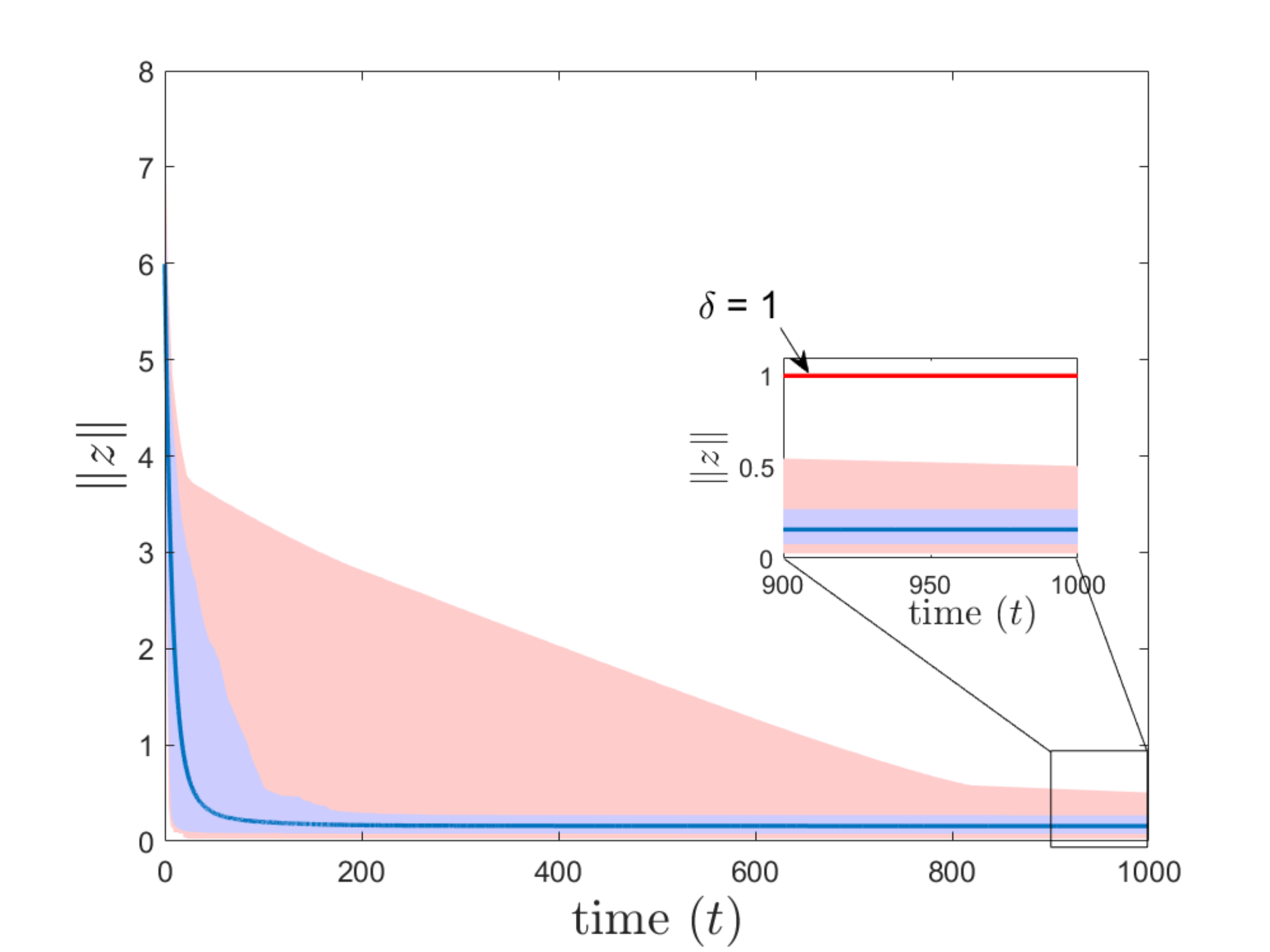}
    \caption{Statistics of the norm of consensus error function $z$ with 95\% confidence interval (blue area) and 100\% data (red area).}
    \label{fig:consensus_statistics}
\end{figure}

\begin{figure}[hbtp!]
    \centering
    \includegraphics[width=0.95\linewidth]{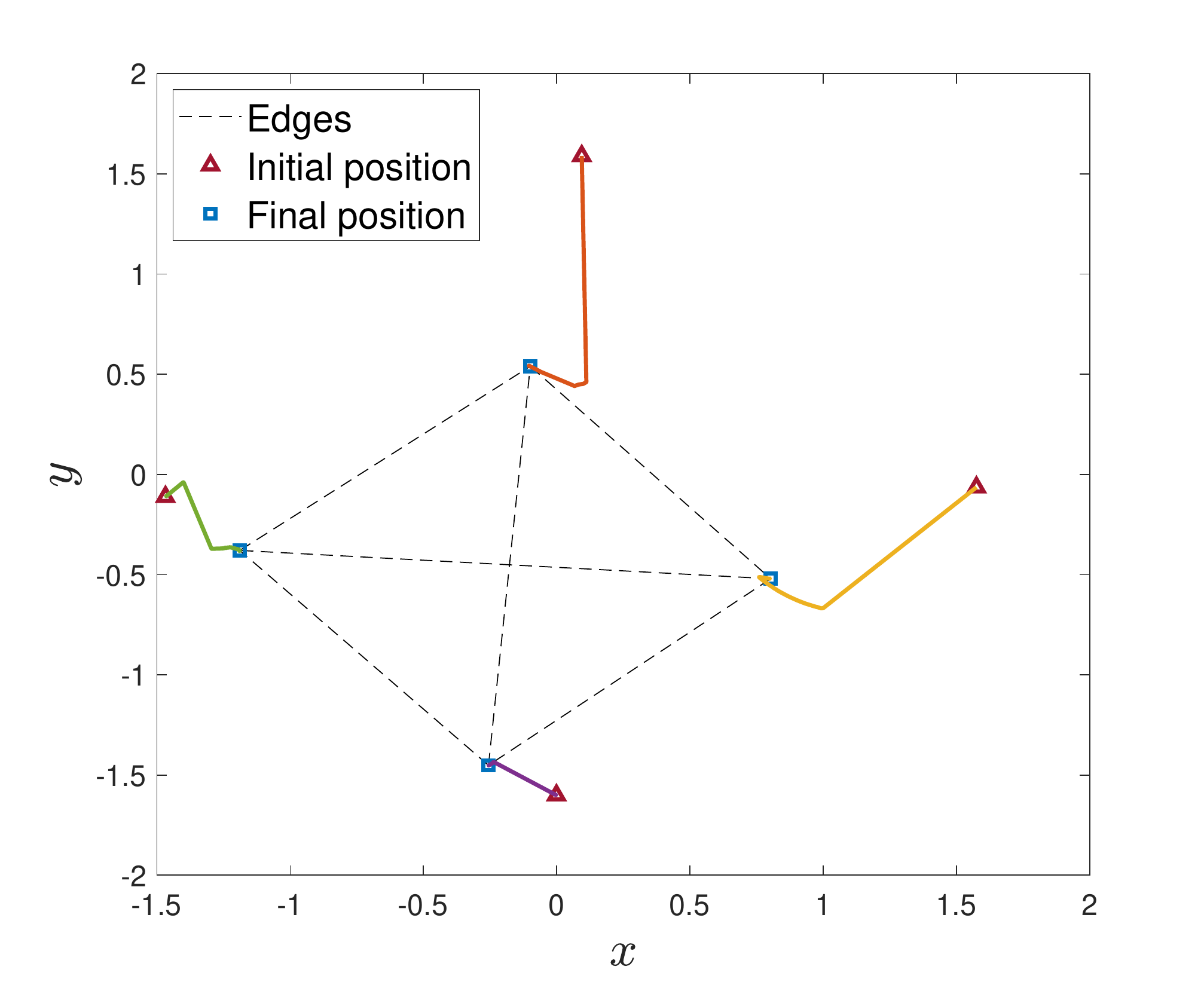}
    \caption{An example of agent trajectories for nearest-neighbor formation control taken from the 1000 random simulations.}
    \label{fig:formation_pos_evol}
\end{figure}

\begin{figure}[hbtp!]
    \centering
    \includegraphics[width=0.95\linewidth]{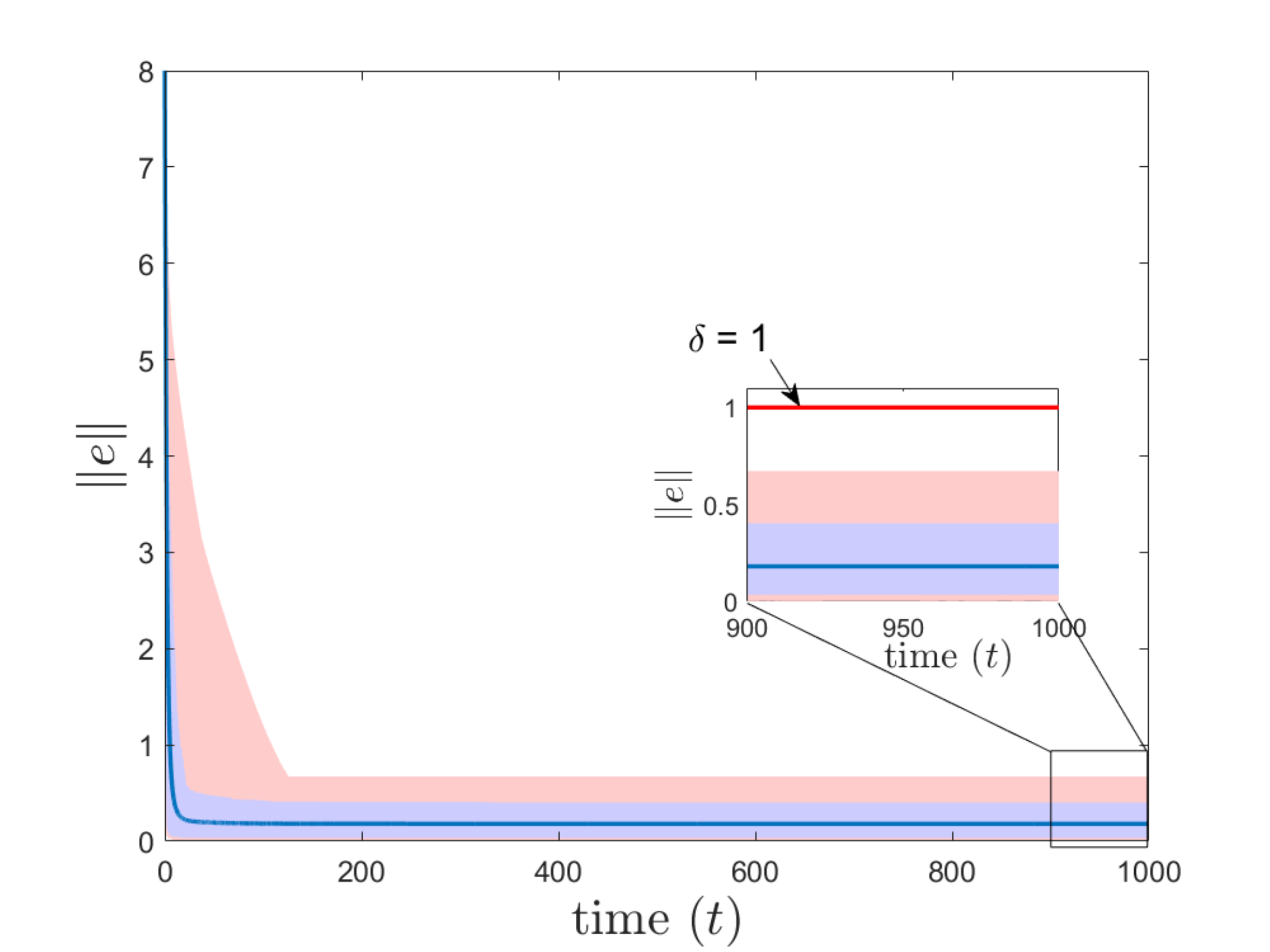}
    \caption{Statistics of the norm of formation error function $e$ with 95\% confidence interval (blue area) and 100\% data (red area).}
    \label{fig:formation_statistics}
\end{figure}

Using the above simulation setup, the results are summarized and presented in Figures \ref{fig:consensus_pos_evol}--\ref{fig:formation_statistics}. 
{\revised The motion animation of both cases can be seen in the following \href{https://youtu.be/ElKByfiTyXY}{video} \url{https://s.id/MAS-NNC}.}
It can be seen from Figure~\ref{fig:consensus_pos_evol} that by using the nearest-neighbor consensus control as proposed in Proposition~\ref{prop:consensus}, the agents reach practical consensus as expected. Furthermore, Fig.~\ref{fig:consensus_statistics} shows that 
in the steady-state, the norm of the error vector $z$ is always below 1 for all samples, which confirms the theoretical result in Proposition \ref{prop:consensus}.

Similar to the consensus case, the nearest-neighbor distance-based formation control as proposed in Proposition~\ref{prop:formation} also performs as expected. In the formation control case, the desired distances between communicating agents are set so that the positions of all agents are on a circle with the radius of $1$. To show the behaviour of the closed-loop systems using the proposed nearest-neighbor distributed control, 
a simulation result of a multi-agent system with four agents (taken from the 1000 random simulations) is shown in Fig.~\ref{fig:formation_pos_evol}. In this plot, all agents converge close to the desired formation shape. The statistical plot of Monte Carlo simulations as given in Fig.~\ref{fig:formation_statistics} shows that the norm of the formation error vector converges to a ball that is smaller than the upper bound as computed in Proposition \ref{prop:formation}.  
This means that all agents converge close to desired formation shape for all simulations. 

{Notably, we can observe from the statistical plots
in Fig.~\ref{fig:consensus_statistics} and Fig.~\ref{fig:formation_statistics} that 
there should be much tighter bounds to the practical stability results as the bounds obtained from the Monte Carlo simulations
is significantly below of the computed bound from Propositions \ref{prop:consensus} and \ref{prop:formation}.}

\section{Conclusion}\label{sec:conclusion}

In this letter, we proposed a nearest-neighbor-based input-quantization procedure for multi agent coordination, namely consensus and distance-based formation control problems where agents can only realize finite set of control points. We have provided rigorous analysis for our proposal. Monte Carlo numerical simulations are presented that confirm the practical stability analysis of both consensus and formation control problems.

\bibliographystyle{IEEEtran}
\bibliography{IEEEabrv,bib_all}

\end{document}